\documentclass[8.5pt,twoside,twocolumn]{article}
\usepackage{amsmath}
\usepackage{amssymb}
\usepackage[version=3]{mhchem}

\usepackage[super,sort&compress,comma]{natbib} 
\usepackage{mhchem}
\usepackage{times,mathptmx}
\usepackage{sectsty}
\usepackage{balance} 

\usepackage{graphicx} 
\usepackage{lastpage}
\usepackage[format=plain,justification=raggedright,singlelinecheck=false,font=small,labelfont=bf,labelsep=space]{caption} 
\usepackage{fancyhdr}
\newcommand{\scs}{\scriptscriptstyle}

\newcommand{\de}{\partial}
\def\<{\langle}
\def\>{\rangle}
\makeatletter
\newcommand*{\rightharpoonupfill@}{%
  \arrowfill@\relbar\relbar\rightharpoonup
}

\newcommand*{\leftharpoondownfill@}{%
  \arrowfill@\leftharpoondown\relbar\relbar
}

\newcommand{\xrightleftharpoons}[2][]{%
  \ensuremath{%
    \mathrel{%
      \settoheight{\dimen@}{\raise 2pt\hbox{$\rightharpoonup$}}%
      \setlength{\dimen@}{-\dimen@}%
      \edef\CA@temp{\the\dimen@}%
      \settoheight\dimen@{$\rightleftharpoons$}%
      \addtolength{\dimen@}{\CA@temp}%
      \raisebox{\dimen@}{%
        \rlap{%
          \raisebox{2pt}{%
            $%
            \ext@arrow 0359\rightharpoonupfill@{\hphantom{#1}}{#2}%
            $%
          }%
        }%
        \hbox{%
          $%
          \ext@arrow 3095\leftharpoondownfill@{#1}{\hphantom{#2}}%
          $%
        }%
      }%
    }%
  }%
}
\makeatother

\begin{document}

\thispagestyle{plain}
\fancypagestyle{plain}{
\fancyhead[L]{\includegraphics[height=8pt]{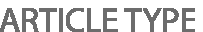}}
\fancyhead[C]{\hspace{-1cm}\includegraphics[height=20pt]{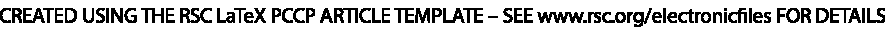}}
\fancyhead[R]{\includegraphics[height=10pt]{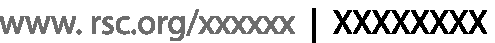}\vspace{-0.2cm}}
\renewcommand{\headrulewidth}{1pt}}
\renewcommand{\thefootnote}{\fnsymbol{footnote}}
\renewcommand\footnoterule{\vspace*{1pt}%
\hrule width 3.4in height 0.4pt \vspace*{5pt}} 
\setcounter{secnumdepth}{5}

\makeatletter 
\def\subsubsection{\@startsection{subsubsection}{3}{10pt}{-1.25ex plus -1ex minus -.1ex}{0ex plus 0ex}{\normalsize\bf}} 
\def\paragraph{\@startsection{paragraph}{4}{10pt}{-1.25ex plus -1ex minus -.1ex}{0ex plus 0ex}{\normalsize\textit}} 
\renewcommand\@biblabel[1]{#1}            
\renewcommand\@makefntext[1]%
{\noindent\makebox[0pt][r]{\@thefnmark\,}#1}
\makeatother 
\renewcommand{\figurename}{\small{Fig.}~}
\sectionfont{\large}
\subsectionfont{\normalsize} 

\fancyfoot{}
\fancyfoot[LO,RE]{\vspace{-7pt}\includegraphics[height=9pt]{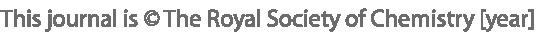}}
\fancyfoot[CO]{\vspace{-7.2pt}\hspace{12.2cm}\includegraphics{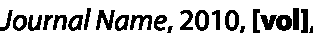}}
\fancyfoot[CE]{\vspace{-7.5pt}\hspace{-13.5cm}\includegraphics{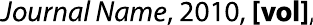}}
\fancyfoot[RO]{\footnotesize{\sffamily{1--\pageref{LastPage} ~\textbar  \hspace{2pt}\thepage}}}
\fancyfoot[LE]{\footnotesize{\sffamily{\thepage~\textbar\hspace{3.45cm} 1--\pageref{LastPage}}}}
\fancyhead{}
\renewcommand{\headrulewidth}{1pt} 
\renewcommand{\footrulewidth}{1pt}
\setlength{\arrayrulewidth}{1pt}
\setlength{\columnsep}{6.5mm}
\setlength\bibsep{1pt}

\twocolumn[
  \begin{@twocolumnfalse}
\noindent\LARGE{\textbf{Diffusion-influenced reactions in a hollow nano-reactor with a circular hole}}
\vspace{0.6cm}

\noindent\large{\textbf{Francesco Piazza,$^{\ast}$\textit{$^{a}$} and Sergey Traytak\textit{$^{b,c}$}}}\vspace{0.5cm}

\noindent\textit{\small{\textbf{Received Xth XXXXXXXXXX 20XX, Accepted Xth XXXXXXXXX 20XX\newline
First published on the web Xth XXXXXXXXXX 200X}}}

\noindent \textbf{\small{DOI: 10.1039/b000000x}}
\vspace{0.6cm}

\noindent \normalsize{
Hollow nanostructures are paid increasing attention in many nanotechnology-related 
communities in view of their numerous applications in chemistry and biotechnology, 
{\em e.g.} as smart nanoreactors or drug-delivery systems.
In this  paper we consider irreversible, diffusion-influenced reactions occurring 
within a hollow spherical cavity endowed with  a circular hole on its surface. 
Importantly, our model is not limited to small sizes of the aperture. 
In our scheme, reactants can freely diffuse inside and outside the cavity through the hole, 
and react at a spherical boundary of given size encapsulated in the chamber and endowed 
with a given intrinsic rate constant.
We work out the solution of the above problem, enabling one to compute 
the reaction rate constant to any desired accuracy. Remarkably, we show that, in the
case of narrow holes, the rate constant is extremely well-approximated by a simple formula 
that can be derived on the basis of simple physical arguments and that
can be readily employed to analyze experimental data.
}
\vspace{0.5cm}
 \end{@twocolumnfalse}
]



\footnotetext{\textit{$^{a}$~Universit\'e d'Orl\'eans, Ch\^ateau de la Source, 45100, Orl\'eans, France,\\
                       Centre de Biophysique Mol\'eculaire, CNRS-UPR4301,
                       Rue C. Sadron, 45071, Orl\'eans, France. Fax: +33 238 631517; Tel: +33 238 255653; 
                       E-mail: Francesco.Piazza@cnrs-orleans.fr}}
\footnotetext{\textit{$^{b}$~Le STUDIUM$^{\small \textregistered}$, 3D av. de la Recherche scientifique, 
                       45071, Orl\'eans, France.}}
\footnotetext{\textit{$^{c}$~Semenov Institute of Chemical Physics RAS,4 Kosygina St.,117977 Moscow, Russia.}}                       


\section{Introduction}
\noindent Chemical processes at the nano-scale are central to many complex phenomena 
in a wide array of modern nanotechnological applications. For example, 
it has long been known that hollow nanostructures provide some advantages in a
number of applications (fillers, pigments, coatings, catalysts etc.) 
because of their lower density~\cite{Yao:2011hc,Chen:2010ij,Skrabalak:2008oq,Srivastava:2004kl,Sun:2003tg}. 
Furthermore, physical and chemical features of hollow nanostructures can nowadays be
fashioned in a controllable manner for a wide range of sizes, shapes,
materials and structural properties of the shells, including thickness, porosity, and
surface reactivity. As a consequence, increasing attention  has been paid over the last decade 
to the elaboration of different engineering methods for manufacturing hollow
nano-objects of various kinds. \\
\indent Among many different nanostructures, hollow spheres and capsules have 
stimulated great interest because of their potential applications in controlled 
drug delivery systems~\cite{Chen:2010ij,Gelder:2005,Lehmann:2014dq}, 
artificial cells~\cite{Li:2002bs}, catalysis~\cite{Chang:2014fk,KimDS:2010}, lithium batteries~\cite{Yao:2011hc} 
and as compartments for confined reactions~\cite{Yang:2010,KimDS:2010,Chattopadhyay:2013fk}. \\
\indent It is clear that many important physical and chemical processes such as
diffusion transfer and chemical reactions might be considerably influenced
by spatial restrictions~\cite{Wu:2012nx,Lu:2011bh,Pich:2011cr,Talkner:2009,Benichou:2010,Konkoli:2009kx}.
Hollow nanostructures find specific applications relative to their bulk counterparts mostly
due to  pronounced size-dependent effects emerging from the confined geometry of
the reaction volumes. Nanochemical processes occurring in confined geometries
usually take place within nano-scale reaction compartments (often referred to as {\em nanoreactors}), 
whose typical dimensions are greater than the relevant reactants sizes~\cite{Feldmann:2010,Fan:2007}.
For example, typical nanoreactors for drug delivery consist of hollow
spheres with reflecting walls and encapsulated
prodrug particles that are needed for the local production of the
appropriate drug. The spherical shells of such nanoreactors have one or several holes 
allowing small particles, reacting with prodrugs, to penetrate inside the nanoreactor by passive
diffusion~\cite{Gelder:2005}. Other kinds of hollow spherical yolk-shell nanoparticles sinthetized as 
delivery vehicles or nanoreactors rely on hierarchical porous structures~\cite{Liu:2010fv}
or are engineered as thermosensitive nano-catalysts~\cite{Wu:2012nx}.\\
\indent Typical dimensions of reactants and compartments ensure that 
reactions occurring in hollow nanostructures and mesoporous materials are mostly
diffusion-influenced. This kind of reactions play an important role
in chemistry and biology, and appropriate mathematical theories are  well 
established for reactions occurring in for unbounded domains~\cite{Rice:1985vn,Zhou:2010}. 
However, despite their great potential importance in many different applications, 
there are very few studies devoted
to diffusion-influenced reactions occurring within hollow spheres. \\
\indent To the best of our knowledge, this problem was first discussed 
by Adam and Delbr\"uck~\cite{Adam:1968ys}. Later Tachiya studied the kinetics of 
diffusion-controlled reactions between particles encapsulated within
a micelle to describe luminescence quenching and excimer formation~\cite{Tachiya:1980}. 
The theory of irreversible, diffusion-influenced quenching reactions 
of the type $A+B^{*} \xrightarrow[]{k} A+B$ occurring at partially
absorbing sinks within a spherical cavity and at the cavity surface were
developed by Bug et al~\cite{Berne:1992}. Three possible schemes for the location of
partially absorbing surfaces within a spherical cavity (acceptors in the center,
at the surface and at both locations) were considered.\\
\indent In another study, the  somewhat similar problem of the desorption of a 
lipid molecule from a lipid vesicle and its incorporation 
into another vesicle at high acceptor concentrations
was reduced to solving the diffusion equation inside  two concentric
spheres~\cite{Almeida:1999zr}. To this end, perfectly absorbing boundary condition were 
imposed on the large sphere and appropriate matching boundary conditions were used on the 
surface of the small sphere. Analogous calculations were performed by L\"u and B\"ulow, who 
solved the diffusion equation in different hollow geometries 
featuring either impermeable or permeable inner cores~\cite{Bulow:2000}. \\
\indent Recently, more sophisticated {\em in silico} schemes based on complex sets of coupled 
reaction-diffusion boundary problems have been introduced with the 
aim of understanding the cellular behavior of toxic foreign compounds.
Methods motivated by homogenization techniques have been applied to make such problem 
treatable, yielding good agreement with experiments~\cite{Berne:2011}.
Along similar lines, the theory of irreversible diffusion-controlled reactions has
been applied to describe reactions between substrates and enzymes in a
whole-cell model~\cite{Vazquez:2010}. However, the Smoluchowski reaction rate constant was used 
in this study, which is questionable when one  takes into account the confined
geometry of the cell and crowding effects.\\
\indent Overall, many studies that investigated reactions within confined geometries 
did not take into consideration the structure of the outer surface, often featuring one of more 
apertures ({\em e.g.} circular pores). Generally speaking, diffusive problems in geometries
of this kind are known as {\em narrow escape} problems~\cite{Schuss:2012}. 
Recently, Sheu and Yang generalized the diffusive narrow escape problem 
to a gating escape model, describing the escape process of a
Brownian particle out of a spherical cavity through a circular gate on the
surface~\cite{Sheu:2010}. The angular size of the aperture was described by a
time-dependent function $\theta _0\left( t\right) $, so that the gate behaves 
like an absorbing or reflecting patch in the open and closed states, respectively. \\
\indent Remarkably, as it is done in Ref.~\cite{Sheu:2010}, 
absorbing boundary conditions are usually imposed on the gate/patch
with the aim of modeling the diffusive escape of a  particle
from a confined volume through a hole on its surface. 
As a consequence,  these theories cannot describe {\em free} diffusion of
particles through the hole. In fact, this would necessitate that the model accommodate 
for both the diffusion from the inside to the exterior and in the opposite direction.
Moreover, the mean first passage time approach is
a powerful tool to study diffusion in compact domains but it is not appropriate for diffusion
in a cavity connected with an outer, unbounded domain.\\
\indent The problem of leakage of Brownian particles through a narrow pore studied in
Ref.~\cite{Singer:2008} is much closer to the problem of free diffusion through a hole, 
as the flux density of the source on the boundary was taken into
account. However, the flux of diffusing particles was given by a prescribed
function and therefore it cannot describe free diffusion of particles
through the hole~\cite{Singer:2008}. Along the same lines, Berezhkovskii and Barzykin studied the
kinetics of diffusive escape from a cavity through a narrow hole in the
cavity wall and successive reentry by a formal kinetic
scheme for reversible dissociation~\cite{Barzykin:2004}.\\
\indent The diffusion-influenced binding to a buried binding site connected to the
surface by a channel studied in Ref.~\cite{Szabo:2011} is the closest problem to the subject of 
our study that can be found in the literaure. Nevertheless, this
problem was only solved for the case of a conical pit with the aid of a constant-flux
approximation or for all geometries where diffusion occurs in 
interior regions that are so narrow that the problem can be approximately considered as
one-dimensional.\\
\indent A thorough analysis of the literature showed that up to now there are no
studies devoted to the theory of diffusion-influenced reactions occurring in hollow spheres
connected through a circular hole to the unbounded outer space containing an excess of
diffusing particles in the bulk. This is the problem that we solve in this paper. \\
\indent The paper is organized as follows. In Section 2 we present a detailed
formulation of the problem at issue. The solution of the problem is described 
in section 3, where we compute the reaction rate constant. In Section 4 we discuss
our results and we show that our problem can be considered as 
equivalent to a much simpler one in the case of very small apertures. The main
conclusions of the paper and possible extensions of the theory are given in
Sec. 5. The appendix contains the details of the calculation and the
explicit expressions of the matrix equations obtained by a dual series
relations approach.

\begin{figure}[t!]
\centering
\includegraphics[width=\columnwidth]{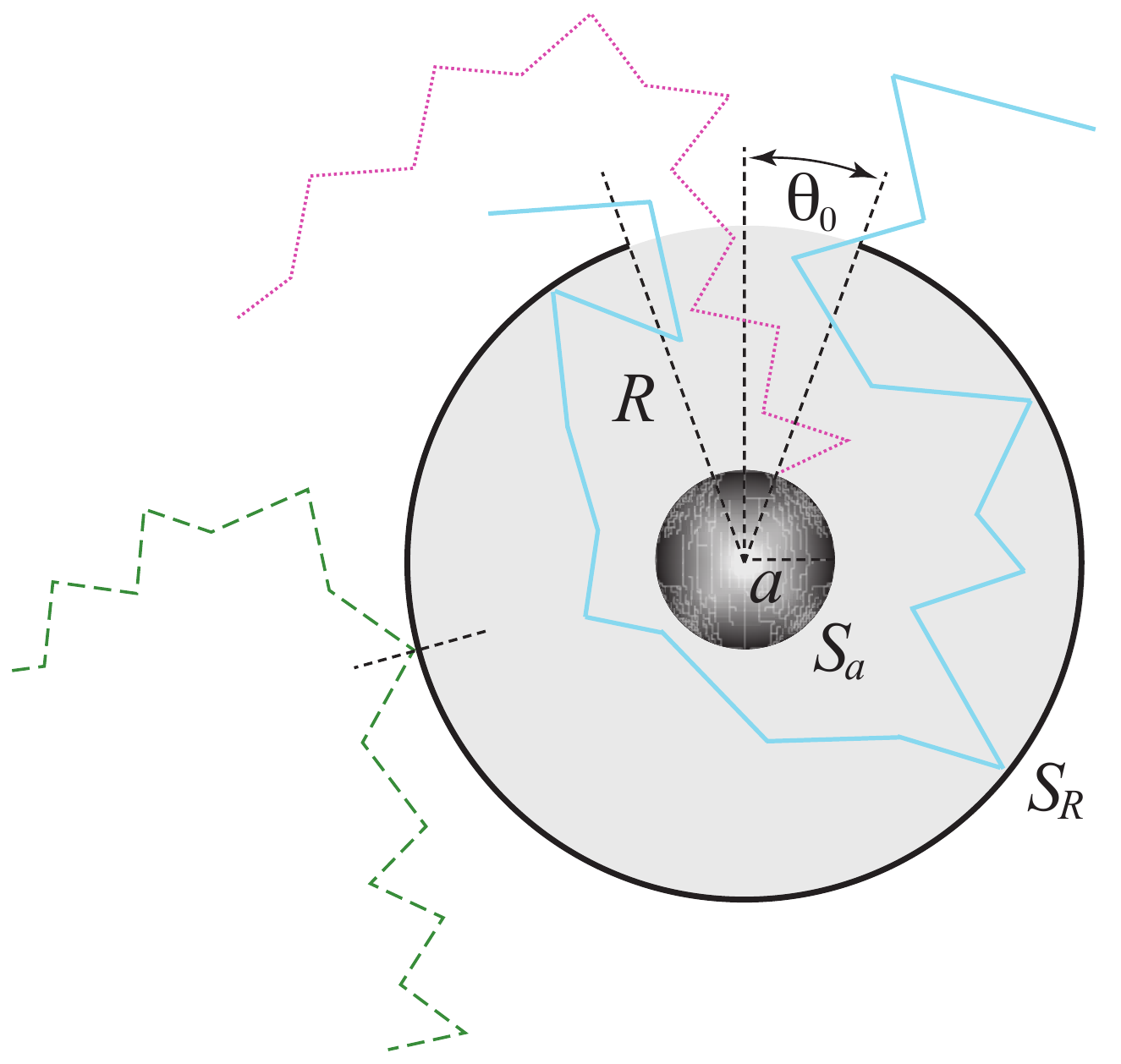}
\caption{(Color online) Schematic representation of our problem. Particles $B$ diffusing from 
the outside (diffusion coefficient $D_{\rm o}$) can be either reflected at the spherical surface $\mathcal{S}_R$
(dashed trajectory) or penetrate through the spherical cap hole. 
In the latter case, they diffuse with coefficient $D_{\rm i}$ and can 
be either absorbed at the inner spherical surface $\mathcal{S}_a$ (dotted trajectory) 
or diffuse back to the exterior through the hole (solid trajectory). \label{f:scheme}}
\end{figure}

\section{The problem}
\noindent Let us consider particles $B$ with bulk density $\rho_{B}$ diffusing into 
a randomly distributed 3D system of hollow spheres with immobile
reactants $A$~(sinks) encapsulated inside them. We assume the hollow spheres to be either fixed in space 
or mobile but much larger than the size of $B$ particles, so that they can be considered as immobile.
For the sake of simplicity, we treat hollow spheres as infinitely thin identical spherical
shells $(S_{R})$ of radius $R$ comprising one spherical sink $(S_{a})$ of radius $a$ ($a\leq R$) and reaction 
surface  $\partial \Omega _a$ each, and featuring an axially symmetric hole $\partial \Omega_0$ 
corresponding to a spherical cap of aperture $\theta_0$ (see Fig.~\ref{f:scheme}).
To make calculations simpler, we consider every sink to be 
concentric with the envelope hollow sphere~\footnote{We note in passing that
this constraint may be removed with the aid of re-expansion formulae
methods~\cite{Traytak:1992,Traytak:1997}. In this way the solution of more general
problems with sinks at arbitrary locations inside the hollow spheres is also feasible.}. 
Introducing a spherical coordinate system $\left( r,\theta ,\varphi \right) $
with the origin at the center of the sphere, we have the following boundaries
\begin{subequations}
\begin{align}
&\partial \Omega _0=\{r=R,0<\theta <\theta _0,0\leq \varphi <2\pi \} \label{e:domW0}\\
&\partial \Omega _1^{\pm }=\{r=R\mp 0,\theta _0<\theta <\pi ,0\leq \varphi<2\pi \} \label{e:domW1} \\
&\partial \Omega _a=\{r=a,0<\theta <\pi,0\leq \varphi <2\pi \}\label{e:domWa}  
\end{align}
\end{subequations}
It is clear that, since the sphere is a two-sided surface, we deal with both the
inside and outside boundaries: $\partial \Omega _1^{+}$ and $\partial \Omega
_1^{-}$ refer to limits taken from within the diffusion subdomains $\Omega
^{+}$ and $\Omega ^{-}$, respectively, where $\Omega ^{+}=\{a<r<R,0<\theta <\pi ,0\leq
\varphi <2\pi \}$ and $\Omega ^{-}=\{R<r,0<\theta <\pi ,0\leq \varphi <2\pi
\}$.\\
\indent In general, keeping in mind, {\em e.g.}, biological applications, we should
consider different media inside and outside the hollow sphere. Thus, we assume that 
the translational diffusion coefficient of particles $B$ can be approximated as
\begin{equation}
D(\mathbf{r})=\left\{ 
\begin{array}{ll}
D_{\rm o} & \mbox{in}\quad \Omega ^{-}\cup \partial \Omega ^{-} \quad (\mbox{outside})\\ 
D_{\rm i} & \mbox{in}\quad \Omega ^{+}\cup \partial \Omega ^{+} \quad (\mbox{inside})
\end{array}
\right. 
\end{equation}
Let us also assume  that the system relaxation time for the diffusive flux of 
$B$ particles $t_D\propto \left( R-a\right) ^2/D_{\rm i}$ is small enough to
neglect time-dependent effects. Hence, in the absence of external forces,
the diffusion of particles $B$ with normalized number density  
$u(\mathbf{r}) =\rho(\mathbf{r})/\rho_{B}$ 
is described by the steady-state diffusion equation
\begin{equation}
\nabla \cdot \left[ D(\mathbf{r})\nabla u(\mathbf{r})\right] =0 \qquad 
\mbox{in}\text{ }\Omega =\Omega ^{+}\cup \Omega ^{-}\cup \partial \Omega _0
\label{sp1}
\end{equation}
which should be solved with the customary bulk boundary condition
\begin{equation}
\label{e:contbulk}
\lim_{|\mathbf{r}|\to\infty} u(\mathbf{r}) = 1
\end{equation}
It is well known from the general theory of partial differential equations
that the classical solution (twice continuously differentiable in $\Omega$
and continuous on $\overline{\Omega}$) of the stationary diffusion equation~\eqref{sp1}
does not exist in the whole domain $\Omega $~\cite{Ladyzhenskaya:1968dz}. 
Therefore one should consider the function 
\begin{equation}
u(\mathbf{r})=\left\{ 
\begin{array}{ll}
u^{-}(\mathbf{r}) & \mbox{for}\quad \Omega^{-} \cup \partial \Omega ^{-}
\\ 
u^{+}(\mathbf{r}) & \mbox{for}\quad \Omega^{+} \cup \partial \Omega ^{+}
\end{array}
\right.\label{sp3}
\end{equation}
Accordingly, continuity of the concentration field across the hole should be enforced, {\em i.e.}
\begin{equation}
       \left. u^{-} \right|_{\partial \Omega_0^{-}}- 
       \left. u^{+} \right|_{\partial \Omega_0^{+}} = 0 \label{e:contu}
\end{equation}
Another condition on the hole can be derived by restricting to 
a small cylinder $\mathcal{C}_{\epsilon}$ of section 
$\delta S \in \partial \Omega_{0}$ with its axis along the normal with height $\epsilon \ll R$, 
{\em i.e.} $\mathcal{C}_{\epsilon}=\{R-\epsilon <r<R+\epsilon ,\left( \theta ,\varphi
\right) \in \delta S\}$.  Using Gauss-Ostrogradsky theorem and Eq.~\eqref{sp1} one has
\begin{equation}
\begin{split}
\lim_{\epsilon \rightarrow 0}\int\limits_{\mathcal{C}_{\epsilon}}\nabla
\cdot \left[ D(\mathbf{r})\nabla u(\mathbf{r})\right] d^3\mathbf{r}=\\
=\int\limits_{\delta S}\left[%
                          D_{\rm i} \frac{\de u^{+}}{\de r} -%
                          D_{\rm o}\frac{\de u^{-}}{\de r}\right]& dS = 0 
\end{split}                          
\end{equation}
Since $\delta S$ is arbitrary, we obtain the following continuity condition
for the local diffusion fluxes, holding at each point of the cap hole $\partial \Omega _0$,
\begin{equation}
\left.\frac{\partial u^-}{\partial r}\right|_{\partial \Omega_0^{-}} = 
\chi \, \left.\frac{\partial u^+}{\partial r}\right|_{\partial \Omega_0^{+}}
\label{e:contf}
\end{equation}
where we have introduced the diffusion anisotropy parameter
\[
\chi = \frac{D_{\rm i}}{D_{\rm o}}
\]
Conditions~\eqref{e:contu} and~\eqref{e:contf} for $D_{\rm i}\neq D_{\rm o}$ 
are often called the {\em weak discontinuity conditions} for the concentration 
field $u(\mathbf{r})$~\footnote{Note that for $D_{\rm i} = D_{\rm o}$ the 
conditions~\eqref{e:contu} and~\eqref{e:contf} turn into continuity conditions 
for $u(\mathbf{r})$.}. To complete the set of boundary conditions, the
two-sided surface of the hollow sphere is assumed to be reflecting both from
the inside $\partial \Omega _1^{+}$ and from the outside $\partial \Omega_1^{-}$, {\em i.e.} 
\begin{equation}
\left. \frac{\de u^{+}}{\de r}\right|_{\partial \Omega_1^{+}} = 
\left. \frac{\de u^{-}}{\de r}\right|_{\partial \Omega_1^{-}} = 0   
\label{e:contref}
\end{equation}
%
%
\subsection{The reaction rate constant}
%

\noindent We are interested in the pseudo-first-order irreversible bulk diffusion-influenced 
reaction between sinks $A$~(encapsulated in hollow spheres with a hole) and reactants $B$ freely 
diffusing in 3D space
\begin{equation}
\label{e:reaction}
A+B 
\xrightleftharpoons[k_{-D}]{k_{D}} 
A\cdot B
\xrightarrow[]{k_{in}}
A+P 
\end{equation}
where $A\cdot B$ denotes the so-called {\em encounter complex}, 
$k_D$ and $k_{-D}$ are the association and dissociation 
diffusive rate constants, respectively, and  $k_{in}$ is the intrinsic 
rate constant of the chemical reaction occurring at the sink surface. 
Reactions of the kind~\eqref{e:reaction} are customary dealt with 
by enforcing radiation boundary conditions~\footnote{This kind of boundary 
conditions are also known as Robin boundary conditions.} at the reaction surface 
$\de \Omega _a$, {\em i.e.} 
\begin{equation}
\left[ 4\pi a^2D_{\rm i} \frac{\de u^{+}}{\de r} - k_{in}u^{+}\right]_{\partial \Omega _a}=0
\label{e:contrad}
\end{equation}
Thus, we can consider that hollow spheres effectively act as sinks of
infinite capacity according to the pseudo-first-order reaction scheme 
\begin{equation}
\label{e:reactionred}
A+B
\xrightarrow[]{k_a}
A+P 
\end{equation}
where the forward diffusion-influenced rate constant $k_a$ is defined by the formula 
\begin{equation}
k_a=\int\limits_{\partial \Omega _a}
                  \left. D_{\rm i} \frac{\de u^{+}}{\de r} \right|_{r=a} dS 
\label{sp9}
\end{equation}
Using this rate constant one can approximately 
describe the kinetics of the effective reaction~\eqref{e:reactionred} as
\begin{equation}
c_B\left( t\right) = c_{B}(0) \exp \left( -k_a c_A t\right)
\label{sp9a}
\end{equation}
where $c_A = const$ is the bulk concentration of hollow spheres,
$c_{B}(t)$ is the time-dependent effective bulk concentration of $B$ particles. 
We stress that our schematization of the problem holds under the {\em excess reactant} 
condition $\rho_{A}\ll \rho_{B}$, $\rho_{A}$ being the bulk number density of sinks. 
Our goal is to compute the rate  constant~\eqref{sp9}.\\
\indent Equation~\eqref{sp1} with the boundary conditions~\eqref{e:contbulk},~\eqref{e:contu},\eqref{e:contf}
and~\eqref{e:contrad} completely specify our mathematical problem. It is expedient in the following to
use the dimensionless spatial variable $\xi =r/R$. The problem at issue can be cast in the following form
\begin{subequations}
\begin{align}
&\nabla^2 u^{\pm} = 0 \quad \mbox{in} \ \Omega^{\pm} \label{e:BVPa}\\
&\left.\frac{\partial u^+}{\partial \xi}\right|_{\xi=\varepsilon} - h u^+(\varepsilon) = 0 
\quad \mbox{for} \ 0 \leq \theta < \pi \label{e:BVPb}\\
&\lim_{\xi\to\infty} u^-(\xi) = 1\label{e:BVPc}\\
&\left.\frac{\partial u^\pm}{\partial \xi}\right|_{\xi=1^{\mp}} = 0 \quad \mbox{for} \
\theta_{0} < \theta < \pi \label{e:BVPd}\\
& \left. u^{+}\right|_{\xi=1^{-}} - \left. u^{-}\right|_{\xi=1^{+}} = 0 
\quad \mbox{for} \ 0 \leq \theta \leq \theta_{0} \label{e:BVPe}\\
&\chi \, \left.\frac{\partial u^+}{\partial \xi}\right|_{\xi=1^{-}} - 
\left.\frac{\partial u^-}{\partial \xi}\right|_{\xi=1^{+}} = 0
\quad \mbox{for} \ 0 \leq \theta \leq \theta_{0} \label{e:BVPf}
\end{align}
\end{subequations}
where $\varepsilon = a/R$ and $h=k_{in} R/(4\pi a^2D_{\rm i})$. The limit~$h\to\infty$ corresponds to 
considering the boundary $\de\Omega_{a}$ as a perfectly absorbing sink. In this case the reaction~\eqref{e:reaction} 
becomes  diffusion-{\em limited}, as the chemical conversion from the encounter complex $A\cdot B$ to the product $P$
becomes infinitely fast with respect to the diffusive step leading to the formation of $A\cdot B$. 

%
\section{The solution}
%
\noindent We look for solutions in the form 
\begin{subequations}
\label{e:solABC}
\begin{align}
&u^{-}(\xi) = 1 + \sum_{n=0}^\infty \frac{A_n}{\xi^{n+1}} \, P_n(\mu) 
\quad \mbox{for} \ \xi \geq 1 \label{e:uoutg}\\
&u^{+}(\xi) = \sum_{n=0}^\infty \left[ \frac{B_n}{\xi^{n+1}}  + C_n \xi^n\right] P_n(\mu)
\quad \mbox{for} \ \xi \leq 1 \label{e:uing}
\end{align}
\end{subequations}
where $A_n,B_n$ and $C_n$ are constants, $\mu = \cos \theta$ and $P_n(\mu)$ are Legendre polynomials of order $n$.
Inserting eq.~\eqref{e:uing} in eq.~\eqref{e:BVPb}, we get 
\begin{equation}
\label{e:BC}
C_n = \alpha_n B_n
\end{equation}
with
\begin{equation}
\label{e:alphan}
\alpha_n = \frac{n+1+h\varepsilon}{\varepsilon^{2n+1}(n-h\varepsilon)}
\end{equation}
Formula~\eqref{sp9} leads to the reduced reaction rate
\begin{equation}
\label{e:flux}
k^{\ast}_{a} = \frac{k_{a}}{k^{+}_{\scs\rm S}} = 
\frac{1}{2} \int_{-1}^{1} \left. \frac{\partial u^{+}}{\partial \xi}\right|_{\xi=\varepsilon}  d \mu
\end{equation}
where $k^{+}_{\scs\rm S} = 4 \pi D_{\rm i} a$ is the {\em internal} Smoluchowski rate
constant for an ideal spherical sink of radius $a$.
Inserting eq.~\eqref{e:uing} in eq.~\eqref{e:flux} and making use of eqs.~\eqref{e:BC} and~\eqref{e:alphan}, we get
\begin{equation}
\label{e:rate}
k_{a}^{\ast} = -\frac{B_0}{\varepsilon}
\end{equation}
So the problem is reduced to the calculation of the constant $B_0$.
The mixed boundary-value problem~\eqref{e:BVPa}-\eqref{e:BVPf} 
can be solved with the method of dual series relations (DSR)~\cite{sneddon:1966}.
DSR admit solutions in the form of an infinite-dimensional system of algebraic equations for a new set of 
unknown coefficients $X_{n},Y_{n}$, that are linearly related to $A_{n},B_{n}$~\cite{Traytak:1995}
\begin{equation}
\label{e:XYsolmain}
Z_n = Z_n^0 + \sum_{m=0}^\infty M_{nm}Z_m,\qquad (n=\overline{0,\infty })
\end{equation}
Here $Z_n=\left( X_n,Y_n\right)^T,Z_n^0=\left( X_n^0,Y_n^0\right)$ and
\begin{equation}
M_{nm}=\left( 
\begin{tabular}{ll}
$M_{nm}^{11}$ & $M_{nm}^{12}$ \\ 
& \\
$M_{nm}^{21}$ & $M_{nm}^{22}$%
\end{tabular}
\right)  
\end{equation}
where $M_{nm}^{ij}$ are four infinite-dimensional matrices of known elements, 
functions of the relevant geometrical and physical parameters $\varepsilon,h$ and $\chi $ 
(see appendix A for the details of the calculation and the explicit expressions of
the matrices $M_{nm}^{ij}$).  In particular, the expression for the rate as a function of 
the new coefficients is
\begin{equation}
\label{e:rateXY}
k_{a}^{\ast} = -\frac{Y_0}\varepsilon 
\end{equation}

%
\section{Results and discussion}
%
%
%
\begin{figure*}[t!]
\centering
\begin{tabular}{c c}
\resizebox{8.2 truecm}{!}{\includegraphics[clip]{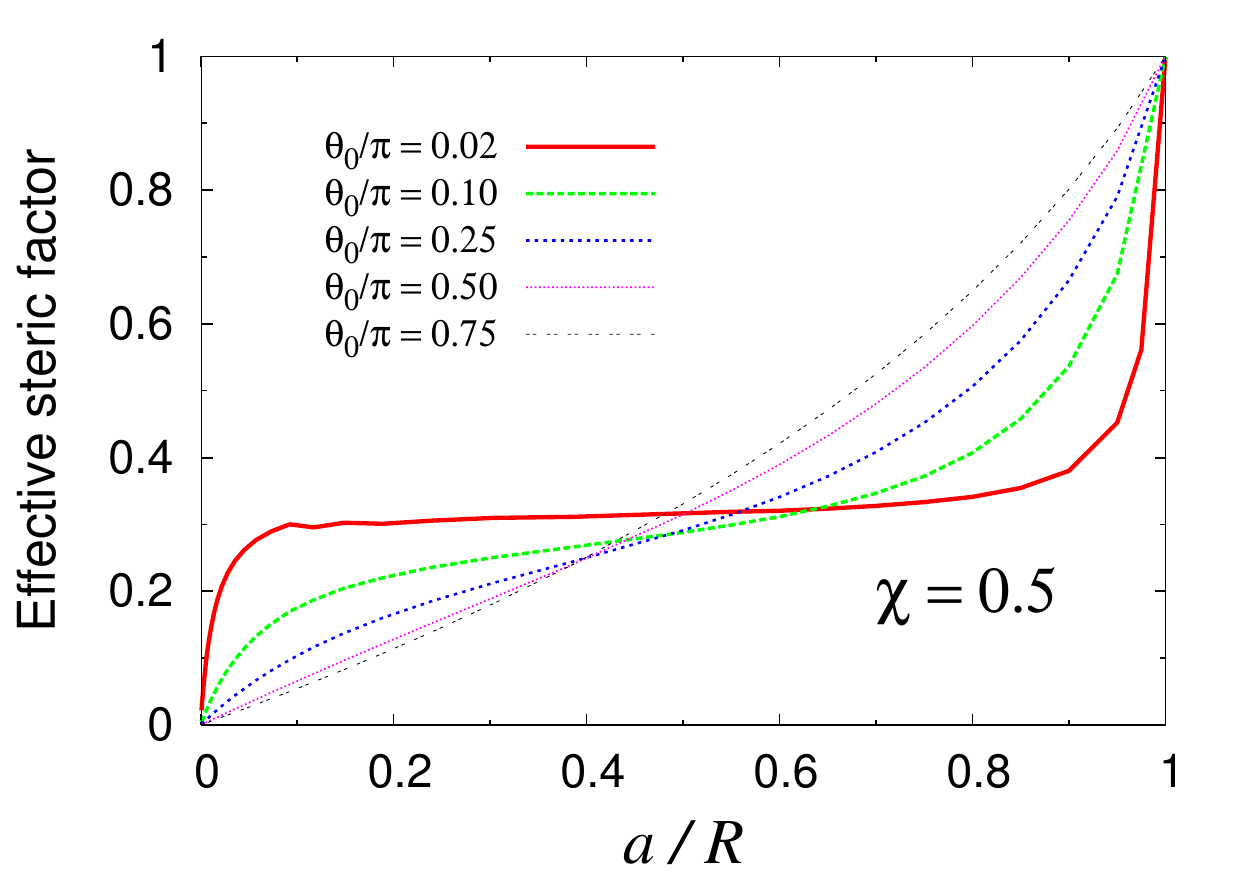}} &
\resizebox{8.2 truecm}{!}{\includegraphics[clip]{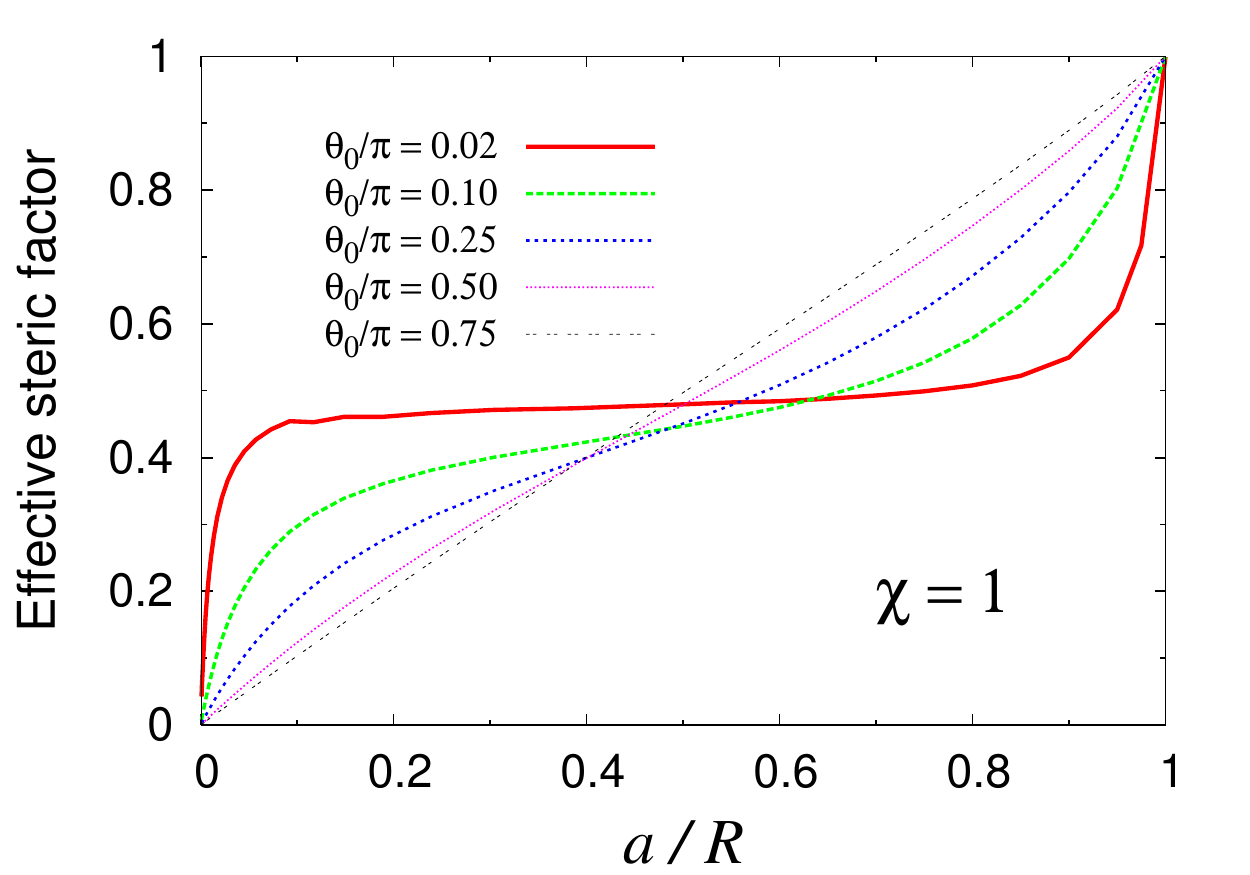}} \\
\resizebox{8.2 truecm}{!}{\includegraphics[clip]{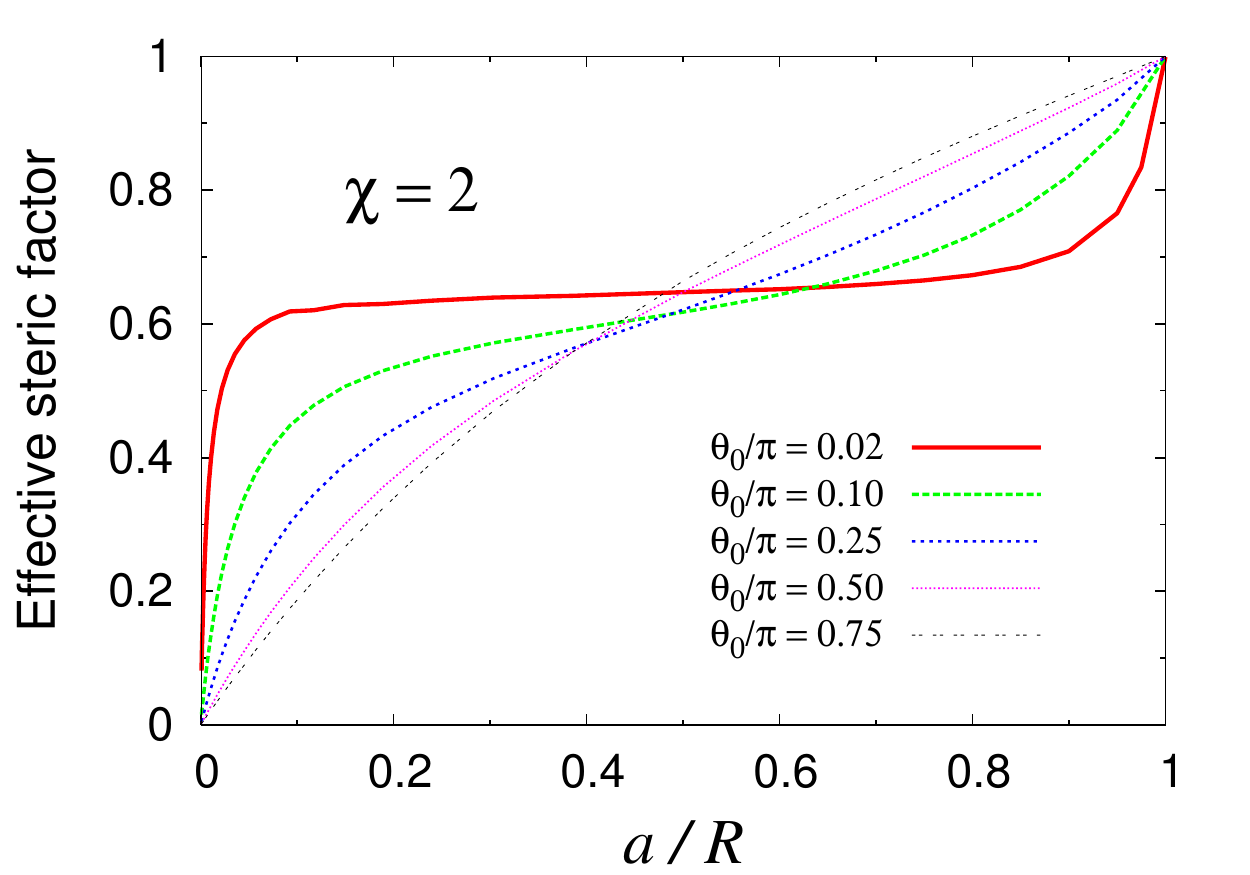}} &
\resizebox{8.2 truecm}{!}{\includegraphics[clip]{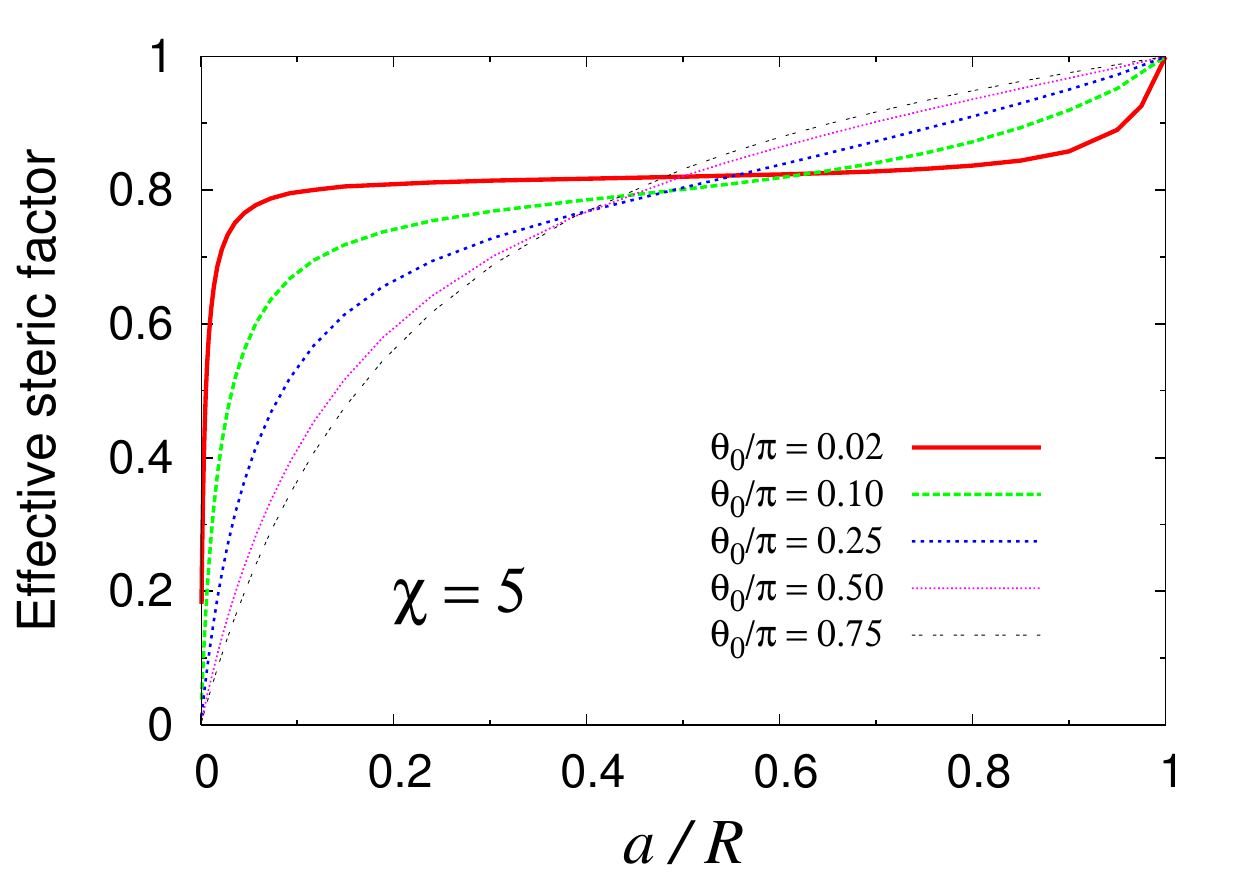}}
\end{tabular}
\caption{\label{f:effsterf}(Color online) Plot of the normalized effective steric 
factor~\eqref{e:effsterfact} as a function of the radius of the inner sphere for 
different sizes of the spherical cap hole and different values of $\chi$. 
The inner encapsulated sphere is taken as perfectly absorbing, {\em i.e.} 
the calculations are performed for $h \to \infty$.}
\end{figure*}
%
\noindent Let us start by considering the limit $\varepsilon\to 1$, that is, $a \to R$.
Furthermore, for the sake of simplicity, let us consider diffusion-limited reactions, 
{\em i.e.} $h\to \infty$.
This case corresponds to considering a perfectly absorbing circular patch 
on an otherwise reflecting sphere of radius $R$. The rate constant for this system can be 
characterized by a steric factor $f_R(\theta_0)\in[0,1]$
\begin{equation}
\label{e:sterfactR}
\frac{k_{\scs R}}{k^{-}_{\scs \rm S}} = f_R(\theta_0) 
\end{equation}
where $k^{-}_{\scs \rm S} = 4 \pi D_{\rm o} R$  is the {\em external} Smoluchowski rate
constant for an ideal spherical sink of radius $R$.
The steric factor $f_R(\theta_0) $ can be calculated to any necessary accuracy 
with the DSR method~\cite{Traytak:1995}. In particular, it was found that 
\begin{equation}
f_R(\theta _0)\sim \frac 1{2\pi }\left( \theta _0+\sin \theta _0\right)
\qquad \mbox{as}\quad \theta _0\rightarrow 0
\label{e:stericf}
\end{equation}
In the general case $a<R$, it is expedient to normalize the reaction rate constant $k_{a}$ to the 
rate constant~\eqref{e:sterfactR}. This is tantamount to 
characterizing the sink inside the spherical cavity 
through a normalized {\em effective} steric factor $\hat{f}(\theta_0;\varepsilon,\chi)\in[0,1]$, 
defined as
\begin{equation}
\label{e:effsterfact}
\begin{split}
\hat{f}(\theta_0;\varepsilon,\chi) 
&:= \frac{k_{a}}{k_{\scs R}} = \frac{k_{a}}{k^{-}_{\scs \rm S} f_R(\theta_0)} \\
&= \frac{\varepsilon\chi}{f_R(\theta_0)}  
\left( \frac{k_{a}}{k^{+}_{\scs\rm S}} \right)
\end{split}
\end{equation}
The physical meaning of $\hat{f}(\theta_{0};\varepsilon,\chi)$ 
is to gauge how {\em effective} is the inner sink of 
radius $a$ in trapping a particle diffusing through the spherical cap hole with respect 
to the situation when the particle is instantaneously trapped the moment it touches 
the cap from the outside ($a=R$). Indeed, as the sink grows to {\em touch} the internal wall of 
the cavity, one has
\begin{equation}
\label{e:limfaR}
\lim_{\varepsilon\to 1} \hat{f}(\theta_0;\varepsilon,\chi) = 1
\end{equation}
independently of $\chi$, as the inner sphere merges with the outer one. Conversely, 
as the sink shrinks, one has 
\begin{equation}
\label{e:limfa0}
\lim_{\varepsilon\to 0} \hat{f}(\theta_0;\varepsilon,\chi) = 0
\end{equation}
uniformly with respect to $\chi$. In this case, 
the effective steric factor vanishes as there is no sink within the 
spherical cavity $\mathcal{S}_R$.\\
\indent In Fig.~\ref{f:effsterf} we plot the normalized effective steric factor as a function of
the inner sink size $a$ for different values of the angular aperture of the circular hole.
As the aperture decreases, $\hat{f}$ feels less and less the dependence on $a$, which 
appears to be limited to two boundary layers in the vicinity of $a=0$ and $a=R$. 
Between the two boundary layers, the effective steric factor is nearly constant. \\
\indent In view of the Gauss-Ostrogradsky  theorem, this is tantamount to saying that 
for small patches the inner sphere {\em feels} a constant flux on the 
surface $r=R$. Hence, the rate does not depend on the surface used for evaluating the 
integral~\eqref{e:flux}.\\
\indent The value of $\hat{f}$ within the plateau is a measure of how much the whole system 
is less effective in trapping a tracer particle from the exterior with respect 
to the patched sphere $\mathcal{S}_R$. Therefore, it is a measure of the 
portion of incoming particle flux through the hole 
that does not reach the inner sink, {\em e.g.} the flux that escapes back to the
exterior through the aperture in the cavity.\\
\indent Interestingly, we see that the such value increases
when the outside diffusion coefficient decreases with respect to the inside (increasing $\chi$).
Recalling that the single-particle diffusion decreases in crowded environment due 
to the  volume occupied by crowding agents, we conclude that, in order to reach diffusively the inner
target more effectively through the hole, the inner medium should be less densely populated 
than the outside. This can be  rationalized in terms of a reduced escape probability towards 
the exterior due to crowding. A different way to picture this effect is to recall that 
in the limit $\chi \to \infty$ the continuity condition~\eqref{e:BVPf} turns the 
spherical hole into a perfectly reflecting patch from the interior. 
Again, no particles allowed to escape outside the spherical cavity.\\
%
\begin{figure}[t!]
\centering
\resizebox{\columnwidth}{!}{\includegraphics[clip]{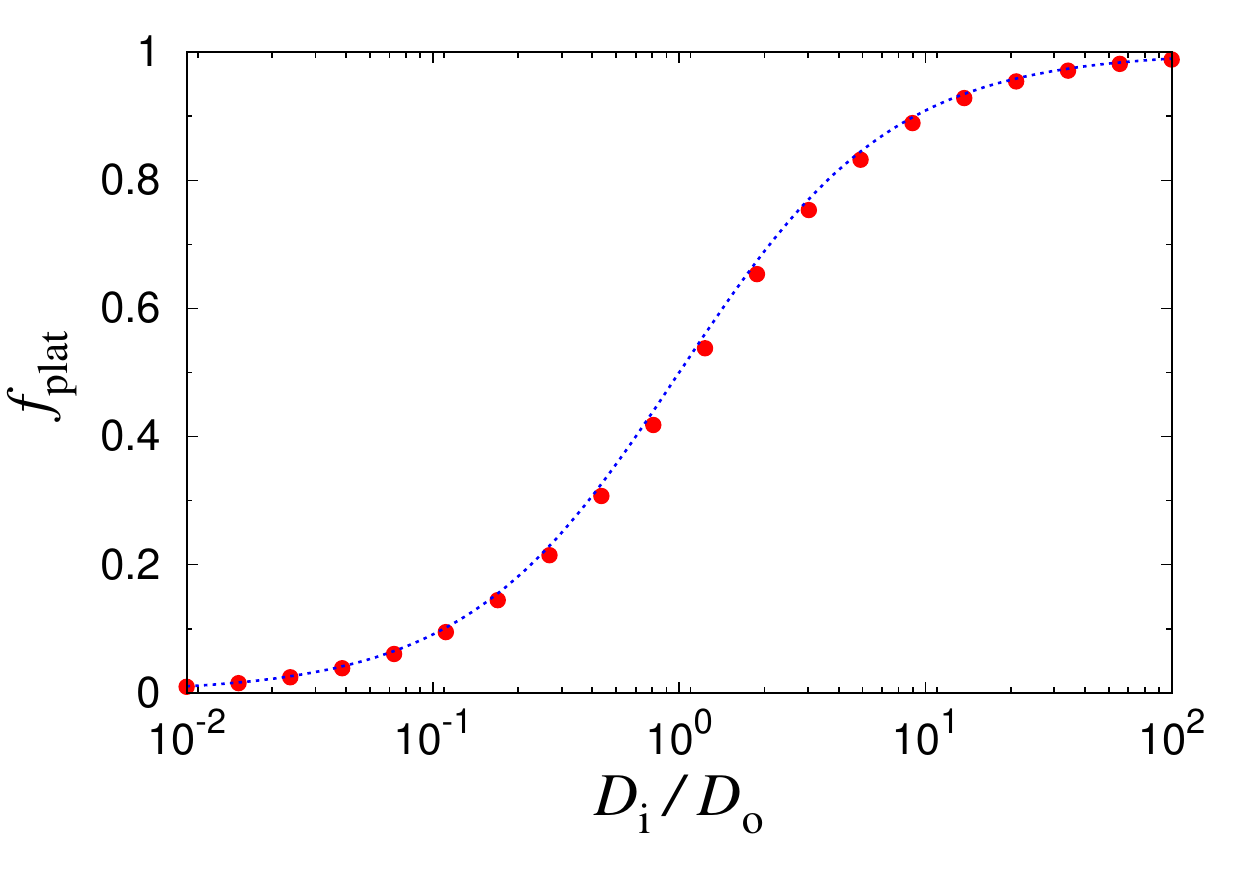}} 
\caption{\label{f:effsterfplat}(Color online) Plot of 
the plateau value of the normalized effective steric factor~\eqref{e:effsterfact},
$\hat{f}_{\rm plat} = \hat{f}(a/R=0.5,\theta_0/\pi=0.02)$
as a function of the inner-to-outer diffusivity ratio $\chi = D_{\rm i}/D_{\rm o}$ (symbols).
The inner sphere is taken as a perfectly absorbing sink, {\em i.e.} the calculation 
is performed for $h \to \infty$. The dotted curve is a plot of the theoretical 
prediction, eq.~\eqref{e:fplateau}.}
\end{figure}
%
%
\indent The plateau value of the effective steric factor for small $\theta_{0}$ 
is proportional to the fraction of flux that reaches the inner sink at equilibrium, 
$\Phi_{\rm in}$, while its complement to one is proportional to the
flux $\Phi_{\rm out}$ that leaves the inside of $\mathcal{S}_R$. 
As $D_{\rm i}/D_{\rm o}$ increases, we see that $\Phi_{\rm out}$
goes to zero, meaning that particles become more and more trapped once they have diffused 
inside $\mathcal{S}_R$. A measure of  $\Phi_{\rm out}$ can thus be obtained 
by plotting the plateau value of $\hat{f}$ as a function of 
$\chi$. This is shown in Fig.~\ref{f:effsterfplat} for the choice $a/R = 0.5$ and $\theta_0/\pi = 0.02$, 
so as to ensure that the boundary layers are sufficiently thin (see again bottom right panel 
in Fig.~\ref{f:effsterf}). We see that, for large values of the ratio $D_{\rm i}/D_{\rm o}$, the 
system behaves as a single sphere of radius $R$ with a small absorbing patch 
(practically no flux leaking back to the outside), {\em i.e.}
$k_{a} \to k^{-}_{\scs \rm S} f_R(\theta_0)$ (see again the definition~\eqref{e:effsterfact}).\\
\indent The flux through the hole that reaches the sink is proportional to $D_{\rm i}$, while the incoming 
flux into the cavity is proportional to $D_{\rm o}$. We can thus surmise that, when the rate into the 
sink becomes independent of its size $a$ for small values of $\theta_{0}$, the fractions of flux
reaching the sink and leaking back through the hole are approximately given by 
$D_{\rm i}/(D_{\rm i} + D_{\rm o})$ and $D_{\rm o}/(D_{\rm i} + D_{\rm o})$, respectively. 
This leads us to conjecture that the value of the effective steric 
factor~\eqref{e:effsterfact} for sinks occupying the 
bulk region of the cavity, {\em i.e.} the plateau shown in Fig.~\ref{f:effsterf}, is given by
\begin{equation}
\label{e:fplateau}
\hat{f}_{\rm plat} = \frac{D_{\rm i}}{D_{\rm i} + D_{\rm o}} = \frac{\chi}{1 + \chi}
\end{equation}
Fig.~\ref{f:effsterfplat} shows that Eq.~\eqref{e:fplateau} yields a perfect 
interpolation of the plateau values, confirming  the validity of our simple 
physical reasoning.
%
%
%
\subsection{Rationalizing the results through a simplified effective model}
%
\noindent  From the above discussion it should be clear that, for small $\theta_0$, we may model our system as 
a single sphere of radius $R$ endowed with a partially absorbing surface, characterized by an 
effective intrinsic reaction rate constant $k_{\rm eff}$. This means that our boundary problem,
for small values of the hole aperture, should become equivalent to the following reduced problem
\begin{subequations}
\begin{align}
&\frac d{d\xi }\left( \xi ^2\frac{du}{d\xi }\right) =0 \label{e:BVPreda}\\
&\left(\frac{\partial u}{\partial \xi} - h_{\rm eff} \, u\right)_{\xi=1} = 0 \label{e:BVPredb}\\
&\lim_{\xi\to\infty} u(\xi) = 1\label{e:BVPredc}
\end{align}
\end{subequations}
The parameter $h_{\rm eff} = k_{\rm eff}/k^{-}_{\scs\rm S}$ gauges the effective {\em absorbing power}
of the sphere $\mathcal{S}_R$. 
This should depend on the steric factor $f_R(\theta_0)$, which guarantees that only a portion of the surface is 
potentially absorbing by construction, and on $\chi = D_{\rm i} /D_{\rm o}$.
It is easy to check that the rate constant $k$ for the above reduced problem is given by  
\begin{equation}
\label{e:solBVPred}
\frac{k}{k^{-}_{\scs\rm S}} = \frac{h_{\rm eff}}{1 + h_{\rm eff}}
\end{equation}
Recalling the definition~\eqref{e:effsterfact}, we see that eqs.~\eqref{e:solBVPred} 
and~\eqref{e:fplateau} fix the effective reactivity of the reduced model, {\em i.e.} 
\begin{equation}
\label{e:heff}
h_{\rm eff} = \frac{\chi f_{R}(\theta_{0})}{ 1 + \chi [1-f_{R}(\theta_{0})]}
\end{equation}
%
%
\section{Conclusion and perspectives}
\noindent In this  paper we investigated an irreversible,
diffusion-influenced reaction occurring within a spherical cavity endowed with 
a circular hole on its surface. Importantly, our model is not limited to small values of the 
angular aperture $\theta_{0}$ of the hole on the cavity surface. 
In our model, $B$ particles can freely diffuse inside and 
outside the cavity through the hole, and react at a spherical boundary $A$ encapsulated 
in the cavity and endowed with a given intrinsic rate constant. This model is relevant 
for chemical and biochemical reactions occurring in hollow nano-structures,
which are intensively studied for a wide array of nanotechnological applications.\\
\indent We work out the solution of the above problem, enabling one to compute 
the reaction rate constant for the encapsulated sphere within the cavity to any 
necessary accuracy.\\
\indent Remarkably, we find that, for small values of the hole aperture,
the rate constant $k_{a}$ becomes independent of the size of the inner reactive sphere.
In this case, the rate is simply proportional to the fraction of diffusive flux that 
is actually absorbed by the sink and thus does not leak back through the hole into the bulk.
We show how this situation can be encapsulated in a simple effective model, 
whose theoretical prediction provides a simple yet powerful formula, {\em i.e.}  
\begin{equation}
\label{e:centralr}
k_{a} = k^{-}_{\scs \rm S} f_{R}(\theta_{0}) \frac{\chi}{1+\chi}
\end{equation}
Here $\chi=D_{\rm i}/D_{\rm o}$ is the ratio of the inside to outside diffusion 
coefficients, $k^{-}_{\scs \rm S} = 4\pi D_{\rm o} R$ is the outside Smoluchowski rate 
constant into the spherical cavity and $f_{R}(\theta_{0}) \simeq (\theta_{0} + \sin \theta_{0})/(2\pi)$ 
is the steric factor that characterizes the rate into the cavity when the hollow sphere 
is modeled as a perfectly absorbing patch (the hole) on an otherwise reflecting surface. 
Eq.~\eqref{e:centralr} is a key result of this paper.\\
\indent Future follow-ups of this work may include extending our mathematical framework to 
diffusion-influenced reactions with two axially symmetric hollow spheres and to 
situations where the encapsulated sink is no longer concentric with the hollow sphere 
but lies at an arbitrary location in the interior~\cite{Vazquez:2011fu}.

%
\section*{acknowledgement}
\noindent This research has been partially supported by Le STUDIUM$^{\small \textregistered}$
(Loire Valley Institute for Advanced Studies), Grant No 2012-109. S. D. T. would like to
thank P. Vigny and N. Fazzalari for their interest in this work.
%
%
%
\appendix
\newpage
%
%
\section{}
\noindent In this appendix we describe in detail the solution of the mixed boundary-value 
problem\eqref{e:BVPa}-\eqref{e:BVPe} with the method of dual series relations (DSR).\\
\indent The constants $A_n$ and $B_n$ can be determined by imposing the boundary conditions~\eqref{e:BVPd} 
and the two continuity conditions~\eqref{e:BVPe} and~\eqref{e:BVPf}. 
Recalling eq.~\eqref{e:BC}, we get the two following coupled DSRs
\begin{subequations}
\begin{align}
&\sum_{n=0}^\infty \left\{
                  (1+\alpha_n)B_n-A_n-\delta_{n0}
                  \right\} P_n(\mu)=0,\nonumber\\   
                  &\qquad\qquad\qquad\qquad\qquad\qquad\qquad  0 \leq \theta \leq \theta_{0} \label{e:DSRAB1}\\
&\sum_{n=0}^\infty (n+1)A_n P_n(\mu)=0,   \quad\quad\quad  \theta_{0} < \theta < \pi \label{e:DSRAB2}\\
&\sum_{n=0}^\infty \left\{
                  (n+1)A_n - \chi [ n(1-\alpha_n) + 1]B_n
                  \right\}P_n(\mu)=0,\nonumber\\     
                  &\qquad\qquad\qquad\qquad\qquad\qquad\qquad\quad  0 \leq \theta \leq \theta_{0}  \label{e:DSRAB3} \\
&\sum_{n=0}^\infty [ n(1-\alpha_n) + 1]B_n
                   P_n(\mu)=0,           \quad \theta_{0} < \theta < \pi \label{e:DSRAB4} 
\end{align}
\end{subequations}
where $\delta_{ij}$ is the Kronecker delta. The above DSRs can be cast in canonical form by 
defining 
\begin{subequations}
\begin{align}
&X_n = \left( \frac{n+1}{2n+1} \right)  A_{n} \label{e:Xn}\\
&Y_n = \left( \frac{n(1-\alpha_n) + 1}{2n+1} \right) B_n\label{e:Yn}
\end{align}
\end{subequations}
which gives
\begin{subequations}
\begin{align}
&\sum_{n=0}^\infty X_{n} P_n(\mu)=G(\theta)    & \qquad 0 \leq \theta \leq \theta_{0}\label{e:DSRXY1} \\
&\sum_{n=0}^\infty (2n+1) X_n P_n(\mu)=0       & \qquad \theta_{0} < \theta < \pi\label{e:DSRXY2} \\
&\sum_{n=0}^\infty Y_{n} P_n(\mu)=F(\theta)    & \qquad 0 \leq \theta \leq \theta_{0}\label{e:DSRXY3} \\
&\sum_{n=0}^\infty (2n+1) Y_n P_n(\mu)=0       & \qquad \theta_{0} < \theta < \pi\label{e:DSRXY4} 
\end{align}
\end{subequations}
with 
\begin{subequations}
\begin{align}
&G(\theta) = \sum_{m=0}^\infty \left[
                                       \frac{X_{m}}{2(m+1)} + \beta_{m} Y_{m}
                                     \right]P_{m}(\cos\theta) -\frac{1}{2} \label{e:GX}\\
&F(\theta) = \sum_{m=0}^\infty \left[
                                       \frac{2m+1}{\chi} \,X_{m} -2m \, Y_{m}
                                     \right]P_{m}(\cos\theta) \label{e:GY}
\end{align}
\end{subequations}
and 
\begin{equation}
\label{e:betam}
\beta_{m} = \frac{(1+\alpha_{m})(2m+1)}{2[m(1-\alpha_{m})+1]}
\end{equation}
%
The DSRs~\eqref{e:DSRXY1},\eqref{e:DSRXY2},\eqref{e:DSRXY3},and~\eqref{e:DSRXY4} admit a formal 
solution in the form of the  infinite-dimensional system of algebraic equations~\cite{sneddon:1966}
\begin{subequations}
\begin{align}
& X_{n} = \frac{\sqrt{2}}{\pi} \int_0^{\theta_{0}} du 
         \cos\left[ \left( n + \frac{1}{2}\right) u \right] \frac{d}{du}
         \int_{0}^{u} \frac{G(\theta) \sin\theta\,d\theta}{\sqrt{\cos \theta - \cos u}} \label{e:Xsol}
         \\
&Y_{n} = \frac{\sqrt{2}}{\pi} \int_0^{\theta_{0}} du 
         \cos\left[ \left( n + \frac{1}{2}\right) u \right] \frac{d}{du}
         \int_{0}^{u} \frac{F(\theta) \sin\theta\,d\theta}{\sqrt{\cos \theta - \cos u}} \label{e:Ysol}
\end{align}
\end{subequations}
The integrals appearing in eqs.~\eqref{e:Xsol} and~\eqref{e:Ysol} can be computed explicitly~\cite{Piazza:2005vn},
by noting that~\footnote{I. S. Gradshteyn and I. M. Ryzhik, Tables of Integrals, Series and Products, 
Academic Press, Eq. 7.225.}
\begin{equation}
\label{e:intLegP}
\int_{0}^{u}\frac{P_{m}(\cos \theta) \sin\theta \, d\theta}{\sqrt{\cos\theta - \cos u}} = 
\frac{2 \sqrt{2}}{2m+1} \sin \left[ \left( m+\frac{1}{2}\right)u \right]
\end{equation}
which finally gives
\begin{eqnarray}
X_{n} = \sum_{m=0}^{\infty} \left( M^{11}_{nm} X_{m} + M^{12}_{nm} Y_{m}  \right) + X^{0}_{n}
         \nonumber\\
Y_{n} = \sum_{m=0}^{\infty} \left( M^{21}_{nm} X_{m} + M^{22}_{nm} Y_{m}  \right) + Y^{0}_{n} \nonumber
\end{eqnarray}
where
\begin{subequations}
\begin{align}
& M^{11}_{nm} = \frac{1}{2(m+1)} \, \Phi_{nm},   \quad
  M^{12}_{nm} =  \frac{(1+\alpha_{m})(2m+1)}{2[m(1-\alpha_{m})+1]}\,\Phi_{nm} 
         \nonumber\\
& M^{21}_{nm} = \frac{2m+1}{\chi}\Phi_{nm}, \quad \quad
  M^{22}_{nm} = -2 m \,\Phi_{nm}   
         \nonumber \\
& X^{0}_{n} = -\frac{\Phi_{n0}}{2}  \qquad Y^{0}_{n} =0\nonumber \\
& \Phi_{nm} = \frac{1}{\pi} \left[
                                  \frac{\sin(m+n+1)\theta_{0}}{m+n+1} \right.\nonumber\\
                                  &\qquad\qquad\qquad +
                                  \left.\frac{\sin (m-n)\theta_{0}}{m-n}(1-\delta_{mn}) + \theta_{0}\delta_{mn}
                                \right]\nonumber
\end{align}
\end{subequations}
Note that $\Phi_{nm}=0$ for $\theta_{0}=0$, which gives $Y_{0}=0$. Hence the rate vanishes in this limit, as it should.
The other interesting limit is $\theta_{0}=\pi$, when the larger external sphere  no longer exists.
In this case it is easy to see that $Y_{0}=(\alpha_{0}+1-\chi)^{-1}$. However, in the limit $\theta_{0}=\pi$,
one has to consider $D_{\rm i} =D_{\rm o} =D$, as the separation between the two spatial domains $r<R$ and $r\geq R$ becomes
immaterial. Hence, recalling eqs.~\eqref{e:alphan} and~\eqref{e:rateXY}, we get
\begin{equation}
\label{e:Y0pi}
\frac{k}{k_{\scs\rm S}} = -\frac{1}{\alpha_{0}\varepsilon}=\frac{h\varepsilon}{1 + h\varepsilon} = 
\frac{k_{in}}{k_{\scs\rm S} + k_{in}} 
\end{equation}
where $k_{\scs\rm S} = 4\pi D a$, which is the correct result for a partially absorbing 
sphere with intrinsic reaction rate constant $k_{in}$.
The limit of fully absorbing sphere $k = k_{\scs\rm S}$ is recovered in the limit $k_{in}\to\infty$.




\balance
\footnotesize{

\providecommand*{\mcitethebibliography}{\thebibliography}
\csname @ifundefined\endcsname{endmcitethebibliography}
{\let\endmcitethebibliography\endthebibliography}{}

}

\end{document}